\documentclass[preprint,aps,prc,superscriptaddress,showpacs,floatfix]{revtex4}
\usepackage{graphicx}

\begin{document}

\title{Cascade production in heavy-ion collisions at \textrm{SIS} energies}
\author{Lie-Wen Chen}
\thanks{On leave from Department of Physics, Shanghai Jiao Tong University,
Shanghai 200030, China}
\affiliation{Cyclotron Institute and Physics Department, Texas A\&M University, College
Station, Texas 77843-3366}
\author{Che Ming Ko}
\affiliation{Cyclotron Institute and Physics Department, Texas A\&M University, College
Station, Texas 77843-3366}
\author{Yiharn Tzeng}
\affiliation{Institute of Physics, Academia Sinica, Taipei 11529, Taiwan}
\date{\today }

\begin{abstract}
Production of the doubly strange $\Xi $ baryon in heavy-ion collisions at 
\textrm{SIS} energies is studied in a relativistic transport model that
includes perturbatively the strangeness-exchange reactions $\bar{K}\Lambda
\rightarrow \pi \Xi $ and $\bar{K}\Sigma \rightarrow \pi \Xi $. Taking the
cross sections for these reactions from the predictions of a hadronic model,
we find that the $\Xi $ yield is about $10^{-4}$ in central collisions of $%
^{58}$Ni + $^{58}$Ni at $E/A=1.93$ \textrm{GeV}. The $\Xi $ yield is further
found to be more sensitive to the magnitude of the cross sections for
strangeness-exchange reactions than to the medium effects due to modified
kaon properties. We have also made predictions for $\Xi $ production in
Au+Au collisions at energies from $1$ to $2$ GeV/nucleon.
\end{abstract}

\maketitle

\section{introduction}

Production of the doubly strange baryon $\Xi $ has been studied in heavy ion
collisions at various energies. For ultrarelativistic energies at SPS, the
measured $\Xi $ abundance is significantly enhanced compared to that
expected from initial nucleon-nucleon collisions \cite{sps}. Explanations
for this enhancement include exotic mechanisms due to the formation of the
quark-gluon plasma, topological defects, and color ropes as well as more
conventional processes of strangeness-exchange reactions such as $\bar{K}%
\Lambda \rightarrow \pi \Xi $ and $\bar{K}\Sigma \rightarrow \pi \Xi $ \cite%
{spal01}. In the latter case, the experimental data can, however, only be
explained if a relatively high dense hadronic matter exists in these
collisions. Observations of $\Xi $ production at lower energies at the AGS,
e.g., Au+Au collisions at 6 AGeV \cite{03E895}, have also been reported
recently, and the measured yield is comparable to the predictions from the
transport model based on strangeness-exchange reactions \cite{spal02}.
Again, the hadronic density reached in these collisions is quite large with
maximum density reaching 7-8 times normal nuclear matter density. In all
these cases, the $\Xi $ yield seems to be consistent with that given by the
statistical model \cite{braun}. Although both $\bar{K}$ and $\Lambda (\Sigma
)$ are infrequently produced in heavy ion collisions at lower GSI energies,
i.e., about 1-2 AGeV, which are below the threshold for producing these
particles from nucleon-nucleon collisions in free space, studies of their
production have provided valuable information about the nuclear equation of
state at high densities as well as the in-medium properties of hadrons \cite%
{Ko96,cassing99}. The latter is believed to be related to the partial
restoration of chiral symmetry in hot dense matter.

In the present study, $\Xi$ production in heavy-ion collisions at \textrm{SIS%
} energies, which are below the threshold of about $3.7$ \textrm{GeV} for
its production in nucleon-nucleon collisions in free space is studied in the
framework of a relativistic transport model. We include $\Xi $ production
from the strangeness-exchange reactions $\bar{K}\Lambda \leftrightarrow \pi
\Xi $ and $\bar{K}\Sigma \leftrightarrow \pi \Xi$ with their cross sections
taken from the predictions of a coupled-channel approach based on a flavor
SU(3)-invariant hadronic Lagrangian \cite{chli02}. The $\Xi $ yield from
these collisions thus depends on the cross sections for these reactions as
well as the abundance of $\bar{K}$, $\Lambda$ and $\Sigma$ in the
collisions. Medium effects due to modified kaon properties are also studied
as it is known to affect the production of $\bar{K}$, $\Lambda$, and $\Sigma 
$ in dense matter \cite{Ko96,cassing99,li94,cassing97,gqli97,gqli98}. We
find that the final $\Xi$ yield is not much affected by the medium effects
but is sensitive to the cross sections for the strangeness-exchange
reactions, thus offering the possibility of testing the predicted cross
sections from the hadronic model.

\section{the relativistic transport model}

The transport model used in present study is taken from that of Refs. \cite%
{gqli97,gqli98} based on the relativistic Vlasov-Uehling-Uhlenbeck equation
(RVUU) \cite{rvuu}. In this model, only the nucleon, delta resonance, and
pion are treated explicitly. Besides undergoing elastic and inelastic
two-body scatterings, these particles also propagate in mean-field
potentials. For nucleons, their potential is taken from the effective chiral
Lagrangian of Ref.\cite{fst} instead of the nonlinear Walecka model used in
Ref.\cite{rvuu}. As a result, their motions are given by the Hamilton
equation of motion, 
\begin{equation}
{\frac{d\mathbf{x}}{dt}}={\frac{\mathbf{p}^{\ast }}{E^{\ast }}},\;\;\;{\frac{%
d\mathbf{p}}{dt}}=-\nabla _{x}(E^{\ast }+W_{0}).
\end{equation}%
where $E^{\ast }=\sqrt{\mathbf{p}^{\ast 2}+m^{\ast 2}}$ with $\mathbf{p}%
^{\ast }=\mathbf{p}-\mathbf{W}$ and $m^{\ast }=m-\Phi $, and $\Phi $ and $%
W=(W_{0},\mathbf{W})$ are the scalar and vector mean-field potentials. For
mean-field parameters, we use the set \textrm{T1} that gives a nuclear
matter incompressibility $K_{0}=194$ \textrm{MeV} and a nucleon effective
mass $m^{\ast }/m=0.60$ at normal nuclear matter density $\rho _{0}=0.15$ 
\textrm{fm}$^{3}$. The $\Delta $ resonance is assumed to have the same
mean-field potential as the nucleon, while the mean-field potential for the
pion is neglected.

Kaons together with its partners (hyperons or antikaons) are produced in
this model from pion-baryon and baryon-baryon reactions, i.e., $\pi
B\rightarrow KY$ and $BB\rightarrow BYK$. In Refs.\cite{gqli97,gqli98}, the
cross section for the reaction $\pi B\rightarrow KY$ is taken from the
resonance model of Ref.\cite{fae94}. Here, we use that from an improved
resonance model \cite{tsush99}. The cross section for the reaction $%
BB\rightarrow BYK$ is, on the other hand, taken from the one-boson-exchange
model of Refs. \cite{liko95b,likoc98}. Antikaons in this model are produced
not only from pion-baryon and baryon-baryon reactions, i.e., $\pi
B\rightarrow K\bar{K}B$ and $BB\rightarrow BBK\bar{K}$, but also from the
pion-hyperon reactions $\pi Y\rightarrow \bar{K}N$ \cite{ko}, where $Y$
denotes either $\Lambda $ or $\Sigma $. Their cross sections are taken
either from predictions of the boson-exchange model or from the empirical
values as in Ref. \cite{gqli97}. Annihilation of produced antikaons is
included via the inverse reactions of pion-hyperon reactions, i.e., $\bar{K}%
N\rightarrow \pi Y$, as other absorption reactions involve the rarely
produced kaons and are thus unimportant. However, the annihilation of kaons
is neglected as it has little effect on kaon production \cite{spal00}.
Because of the small production probabilities of kaons, hyperons, and
antikaons in heavy-ion collisions at \textrm{SIS} energies, the above
discussed reactions are treated perturbatively. In this method, the dynamics
of nucleons, $\Delta $ resonances, and pions that produce these particles
are not affected, and the produced particle is given a probability
determined by the ratio of its production cross section to the total
two-body scattering cross section.

After being produced, these rare particles also undergo elastic and
inelastic scattering as well as propagate in mean-field potentials. For $%
\Lambda $ and $\Sigma $, their mean-field potentials are taken to be $2/3$
of the nucleon potential according to their light quark content. For kaon
and antikaon, their mean-field potentials are obtained from the chiral
Lagrange including both scalar and vector interactions \cite{gqli97}, i.e., 
\begin{equation}  \label{kaon}
\omega _{K,\bar{K}}=\left[ m_{K}^{2}+\mathbf{k}^{2}-a_{K}\rho
_{S}+(b_{K}\rho )^{2}\right] ^{1/2}\pm b_{K}\rho .
\end{equation}
In the above, we have $b_{K}=3/(8f_{\pi }^{2})\approx 0.333$ \textrm{GeVfm}$%
^{3}$ corresponding to a pion decay constant $f_{\pi }\approx 103$ MeV; $%
\rho $ and $\rho _{s}$ are, respectively, the nuclear matter and scalar
density; and the ``$+$'' and ``$-$'' signs in the last term are for kaon and
antikaon, respectively. The two parameters $a_{K}$ and $a_{\bar{K}}$, which
determine the strength of the attractive scalar potential for kaon and
antikaon, respectively, can in principle be evaluated from the chiral
Lagrangian \cite{lee}. Here, we follow the method of Ref.\cite{gqli97} by
determining their values from fitting the experimental data on kaon and
antikaon yield in heavy-ion collisions using the RVUU model, and this will
be described in detail later.

The perturbative method for rare particle production in heavy ion collisions
was first introduced in Ref.\cite{ko80} and is now extensively used in
transport models \cite{Ko96,cassing99}. Since the yield of $\Xi $ particle
is expected to be small at SIS energies, it is treated perturbatively as
well. The produced $\Xi $ from the strangeness-exchange reactions $\bar{K}%
Y\rightarrow \pi \Xi $ is thus given a probability equal to the ratio of
this cross section to the $\bar{K}Y$ total cross section, which is taken to
be $20$ \textrm{mb}, multiplied by the probabilities carried by $\bar{K}$
and $Y$. Since the final $\Xi $ yield is given by the product of its
production probability in the reaction $\bar{K}Y\rightarrow \pi \Xi $ and
the number of $\bar{K}Y$ collisions, which is proportional to the $\bar{K}Y$
total cross section, it depends only on the cross section for the reaction $%
\bar{K}Y\rightarrow \pi \Xi $ but not on the $\bar{K}Y$ total cross section.
During its propagation in the nuclear medium, the $\Xi $ is subjected to a
mean-field potential that is $1/3$ of the nucleon potential as it has only
one light quark.

\section{the strangeness-exchange reactions $\bar K\Lambda\to\protect\pi\Xi$
and $\bar K\Sigma\to\protect\pi\Xi$}

\begin{figure}[ht]
\includegraphics[height=3.0in,width=3.2in]{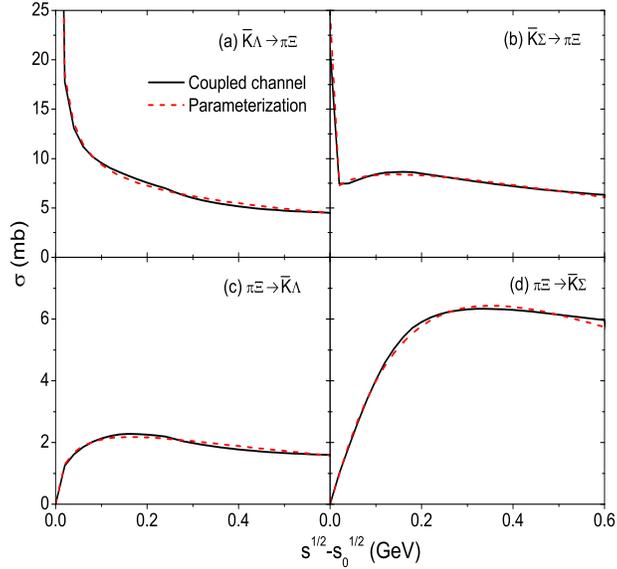}
\caption{Isospin-averaged cross sections for (a) $\bar{K}\Lambda \rightarrow 
\protect\pi \Xi $, (b) $\bar{K}\Sigma \rightarrow \protect\pi \Xi $, (c) $%
\protect\pi \Xi \rightarrow \bar{K}\Lambda $, and (d) $\protect\pi \Xi
\rightarrow \bar{K}\Sigma $. Solid curves are from the coupled-channel
calculations of Ref. \protect\cite{chli02} with a cutoff parameter of 1 GeV,
and dashed curves are parameterized ones based on Eq.(\ref{param}).}
\label{crsc}
\end{figure}

For the strange-exchange reactions $\bar{K}Y\rightarrow \pi \Xi $, there is
no empirical information on their cross sections. These reactions have,
however, been studied in the coupled-channel approach based on a gauged
flavor \textrm{SU(3)}-invariant hadronic Lagrangian \cite{chli02}. The spin-
and isospin-averaged cross sections for these reactions can be written as 
\begin{equation}
\sigma _{\bar{K}\Lambda \rightarrow \pi \Xi }=\frac{1}{4}\frac{p_{\pi }}{p_{%
\bar{K}}}|M_{\bar{K}\Lambda \rightarrow \pi \Xi }|^{2},\qquad \sigma _{\bar{K%
}\Sigma \rightarrow \pi \Xi }=\frac{1}{12}\frac{p_{\pi }}{p_{\bar{K}}}|M_{%
\bar{K}\Sigma \rightarrow \pi \Xi }|^{2},\qquad
\end{equation}%
where $p_{\bar{K}}$ and $p_{\pi }$ are initial antikaon and final pion
momenta in the center-of-mass system, and $|M_{\bar{K}\Lambda \rightarrow
\pi \Xi }|^{2}$ and $|M_{\bar{K}\Sigma \rightarrow \pi \Xi }|^{2}$ are
squared invariant matrix elements with summation over the spins and isospins
of both initial and final particles. The results from the coupled-channel
approach with cutoff parameter of $1$ GeV can be fitted by using the
following parameterization for the squared invariant matrix elements: 
\begin{equation}
|M_{\bar{K}\Lambda \rightarrow \pi \Xi }|^{2}=34.7\frac{s_{0}}{s}~\mathrm{mb}%
,\qquad |M_{\bar{K}\Sigma \rightarrow \pi \Xi |}|^{2}=318\left( 1-\frac{s_{0}%
}{s}\right) ^{0.6}\left( \frac{s_{0}}{s}\right) ^{1.7}~\mathrm{mb}
\label{param}
\end{equation}%
where the threshold energy $s_{0}^{1/2}$ in the center-of mass system is $%
1.611$ \textrm{GeV} and $1.688$ \textrm{GeV} for the reactions $\bar{K}%
\Lambda \rightarrow \pi \Xi $ and $\bar{K}\Sigma \rightarrow \pi \Xi $,
respectively. In Figs. \ref{crsc}(a) and (b), we show by solid curves the
cross sections for the two reactions $\bar{K}\Lambda \rightarrow \pi \Xi $
and $\bar{K}\Sigma \rightarrow \pi \Xi $ from the coupled-channel approach
and by dashed curves those using the parameterizations of Eq.(\ref{param}).
Except near threshold, these cross sections are of the order of $5$-$10$ 
\textrm{mb}.

The cross sections for the inverse reactions $\pi\Xi\to\bar K\Lambda$ and $%
\pi\Xi\to\bar K\Sigma$, which are needed for treating $\Xi$ annihilation,
are related to those for $\Xi$ production by the principle of detailed
balancing, i.e., 
\begin{equation}
\sigma_{\pi\Xi\rightarrow\bar{K}\Lambda}=\frac{1}{3} \frac{p_{\bar{K}}^2}{%
p_{\pi}^2}\sigma_{\bar K\Lambda\to\pi\Xi}, \qquad \sigma_{\pi\Xi\rightarrow%
\bar{K}\Sigma}=\frac{p_{\bar{K}}^2}{p_{\pi}^2} \sigma_{\bar
K\Sigma\to\pi\Xi}.
\end{equation}
These cross sections are shown in Figs. \ref{crsc}(c) and (d) by solid
curves for those from the coupled-channel approach and by dashed curves for
the parameterized ones.

\section{kaon and antikaon dynamics}

\begin{figure}[th]
\includegraphics[height=3.0in,width=3.0in]{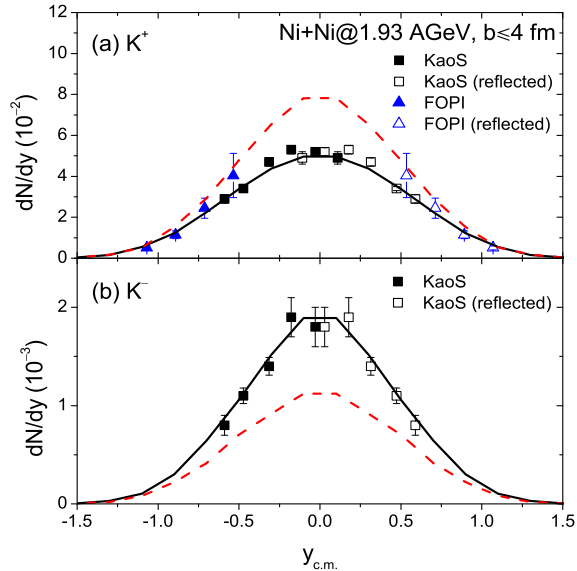}
\caption{Rapidity distributions of (a) $K^{+}$ and (b) $K^{-}$ with (solid
lines) and without (dashed lines) kaon medium effects in near-central
collisions of $^{58}$Ni + $^{58}$Ni at $E/A=1.93$ GeV. The experimental data
are from the FOPI \protect\cite{Best97} and KaoS \protect\cite{Menzel00}
collaborations.}
\label{dndykpkm}
\end{figure}

Since the effective masses of kaons and antikaons are modified in nuclear
medium, their production cross sections differ from those in free space.
Following Refs.\cite{gqli97,gqli98}, their values at center-of-mass energy $%
\sqrt{s^{\ast }}$-$\sqrt{s_{\mathrm{threshold}}^{\ast }}$ above the
threshold energy $\sqrt{s_{\mathrm{threshold}}^{\ast }}$ in the medium are
taken to have same values as those at center-of-mass energy $\sqrt{s}$-$%
\sqrt{s_{\mathrm{threshold}}}$ above the threshold energy $\sqrt{s_{\mathrm{%
threshold}}}$ in free space. Taking into account also propagation of kaons
and antikaons in their mean-field potentials given by Eq.(\ref{kaon}), we
have studied kaon and antikaon production in near-central collisions of $%
^{58}$Ni + $^{58}$Ni at $E/A=1.93$ \textrm{GeV}. We find that the measured
rapidity distributions of kaons and antikaons \cite{Best97,Menzel00}, shown
by squares in Figs. \ref{dndykpkm}(a) and (b), respectively, can be fitted
with the parameters $a_{K}=0.22$ \textrm{GeV}$^{2}$\textrm{fm}$^{3}$ and $a_{%
\bar{K}}=0.35$ \textrm{GeV}$^{2}$\textrm{fm}$^{3}$. These values are
somewhat smaller than those used in Ref. \cite{gqli97prl} since slightly
different cross sections are used in the present study as discussed in the
previous section.

The above parameters give a repulsive potential of about $20$ \textrm{MeV}
for kaons and an attractive potential of about $-98$ \textrm{MeV} for
anitkaons with zero momentum in a nuclear matter at normal density. This
value of kaon potential is consistent with that known empirically from the
kaon-nucleus scattering \cite{barnes} and also from a more recent study of
soft kaon production in p+A reactions at COSY-J\"{u}lich \cite{nekipelov}.
The magnitude of the antikaon potential is smaller than that extracted from
the kaonic atom data \cite{friedman}, which is about $-200\pm 20$ \textrm{MeV%
} but depends strongly on the extrapolation procedure from the surface of
nuclei to their interiors \cite{friedman}. It is, however, comparable to the
moderate attractive potential with magnitude of about $50$-$90$ \textrm{MeV}
predicted by recent studies based on self-consistent calculations with a
chiral Lagrangian \cite{lutz98,ramos00} or meson-exchange potentials \cite%
{tolos01}. The relatively shallow antikaon-nucleus potential has also been
found to reproduce reasonably the experimental data on kaonic atoms in
studies based on the SU(3) chiral unitary model \cite{hirenzaki00}. The
present values of kaon and antikaon potentials determined from fitting the
experimental data on kaon and antikaon yield in heavy-ion collisions using
the RVUU model are thus reasonable.

In Fig. \ref{dndykpkm}, we also show the results from the RVUU model without
kaon medium effects. It is seen that the repulsive kaon potential reduces
kaon production due to the higher threshold while the attractive antikaon
potential enhances antikaon production as a result of lower threshold.
Including medium effects on kaon and antikaon production in heavy ion
collisions are thus important in describing the measured kaon and antikaon
yields.

\section{results}

\subsection{Nonstrange and singly strange hadrons}

\begin{figure}[ht]
\includegraphics[height=4.2in,width=3.2in]{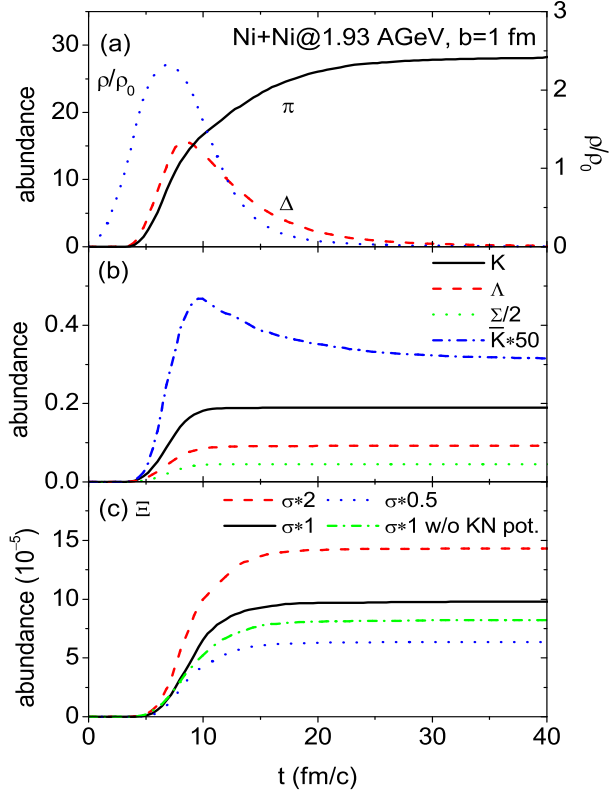}
\caption{Time evolutions of (a) central baryon density (right scales) and
abundances (left scales) of $\protect\pi$, $\Delta$; (b) abundances of $K$, $%
\Lambda $, $\Sigma $, and $\bar{K}$; and (c) the $\Xi$ abundance in cases
with and without kaon medium effects and different cross sections (see text)
in collisions of $^{58}$Ni + $^{58}$Ni at $E/A=1.93$ GeV at $b=1$ fm.}
\label{abndtime}
\end{figure}

We first show in Fig. \ref{abndtime} the time evolutions of (a) central
baryon density and the abundances of $\pi $, $\Delta$, and (b) $K$, $\Lambda 
$, $\Sigma $, $\bar{K}$ in collisions of $^{58}$Ni + $^{58}$Ni at energy $%
E/A=1.93$ \textrm{GeV} and at impact parameter $b=1$ \textrm{fm}. It is seen
that the colliding system reaches the highest compression of central baryon
density of about $2.3\rho _{0}$ at about $7$ \textrm{fm/c}, and most
particles are produced during this compression stage. The final $\pi$
multiplicity is about $28$, corresponding to an averaged charged pion
multiplicity $(\pi ^{+}+\pi ^{-})/2$ of about $9.3$, which is in good
agreement with the experimental data \cite{97fopi}. The abundances of $K$, $%
\Lambda $, and $\Sigma $ reach their respective final values of $0.189$, $%
0.092$, and $0.091$ quickly at about $10$ \textrm{fm/c}, while the $\bar{K}$
abundance first increases and then decreases to its final value of about $%
6.3\times 10^{-3}$ at around $30$ \textrm{fm/c}. The decrease of the $\bar{K}
$ abundance is due to the strong $\bar{K}$ absorption through the
strangeness-exchange reactions $\bar{K}N\rightarrow \pi Y$. The final $\bar{K%
}/K$ ratio is about $3.3\times 10^{-2}$ and is again in good agreement with
experimental data and is also comparable to the predictions from the
statistical model \cite{Oeschler01}.

\subsection{$\Xi$ production from $^{58}$Ni + $^{58}$Ni collisions at $%
E/A=1.93$ GeV}

The time evolution of the $\Xi $ abundance from this reaction is shown by
the solid curve in Fig. \ref{abndtime}(c). The $\Xi $ abundance is seen to
reach its final value of about $9.8\times 10^{-5}$ at about $20$ \textrm{fm/c%
}. The resulting $\Xi /K$ ratio is thus about $5.2\times 10^{-4}$ and is
measurable in experiments at GSI. To see how the $\Xi $ abundance depends on
the cross sections for the strangeness-exchange reactions $\bar{K}\Lambda
\leftrightarrow \pi \Xi $ and $\bar{K}\Sigma \leftrightarrow \pi \Xi $, we
also show in Fig. \ref{abndtime} (c) the time evolution of the $\Xi $
abundance from same heavy-ion collisions when the cross sections for the
strangeness-exchange reactions are varied by a factor $2$. This changes the $%
\Xi $ abundance to $1.4\times 10^{-4}$ and $6.3\times 10^{-5}$,
corresponding to effects of about $46\%$ and $36\%$, when the cross sections
for the strangeness-exchange reactions are doubled (dashed line in Fig. \ref%
{abndtime} (c)) and halved (dotted line in Fig. \ref{abndtime} (c)). Our
results thus show that the $\Xi $ yield in heavy ion collisions does not
reach chemical equilibrium, similar to the conclusion of Ref.\cite{hartnack}
on subthreshold $\bar{K}$ production in heavy-ion collisions at SIS energies.

In Fig. \ref{abndtime}(c), we also show by the dash-dotted curve the $\Xi $
abundance in the case of neglecting kaon medium effects. In this case, the
final $\Xi $ yield is about $8.2\times 10^{-5}$ and is only about $16\%$
less than that from including kaon medium effects. The small effect is due
to the fact that kaon medium effects increase the $\bar{K}$ abundance but
decrease the abundance of $\Lambda $ and $\Sigma $. The $\Xi $ abundance is
thus more sensitive to the cross sections for strangeness-exchange reactions
than to the medium effects due to the modification of kaon properties. Study
of $\Xi $ production in heavy-ion collisions at SIS energies thus provides
the possibility to understand the strangeness-exchange reactions $\bar{K}%
\Lambda \leftrightarrow \pi \Xi $ and $\bar{K}\Sigma \leftrightarrow \pi \Xi 
$, and to test the hadronic model used in Ref.\cite{chli02} for studying
these reactions.

\begin{figure}[th]
\includegraphics[height=3.0in,width=3.0in]{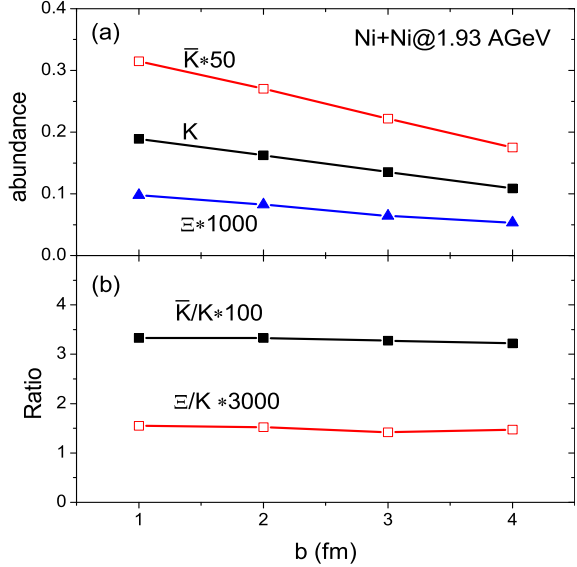}
\caption{(a) Abundances of $K$, $\bar{K}$, and $\Xi$, and (b) $\bar{K}/K$
and $\Xi /K$ ratios as functions of impact parameter $b$ in collisions of $%
^{58}$Ni + $^{58}$Ni at $E/A=1.93$ GeV.}
\label{bdep}
\end{figure}

To see the centrality dependence of particle production, we show in Figs. %
\ref{bdep}(a) the abundances of $K$, $\bar{K}$, and $\Xi $, and (b) the $%
\bar{K}/K$ and $\Xi /K$ ratios as functions of impact parameter $b$ in
collisions of $^{58}$Ni + $^{58}$Ni at $E/A=1.93$ \textrm{GeV}. It is seen
that the abundances of $K$, $\bar{K}$ and $\Xi $ decrease almost linearly
with $b$ while the $\bar{K}/K$ and $\Xi /K$ ratios only depends weakly on
centrality. For the $\bar{K}/K$ ratio, its value is in the range $3.2\times
10^{-2}\sim 3.3\times 10^{-2}$ for impact parameters between 1 and 4 fm and
agrees well with the experimental data \cite{Oeschler01}. The weak
centrality dependence of $\bar{K}/K$ was also observed in previous
calculations \cite{gqli98}. The predicted variation of the $\Xi /K$ ratio
with impact parameter is in the range $4.7\times 10^{-4}\sim 5.2\times
10^{-4}$.

\subsection{$\Xi$ production from $^{197}$Au + $^{197}$Au}

\begin{figure}[th]
\includegraphics[height=3.0in,width=3.0in]{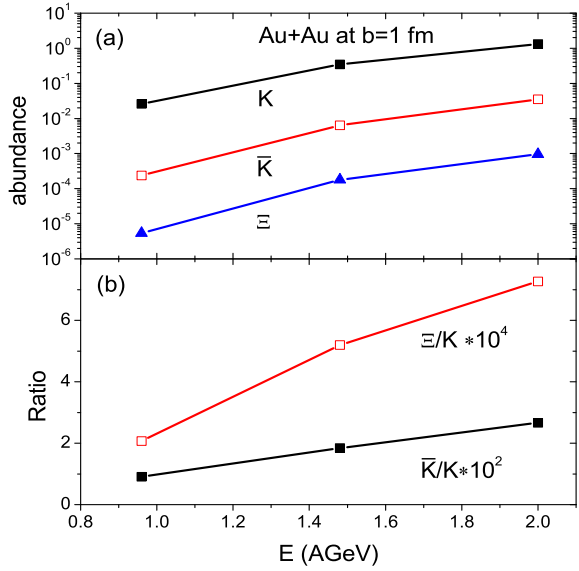}
\caption{Excitation functions for (a) the abundance of $K$ , $\bar{K}$, and $%
\Xi $, and (b) $\bar{K}/K$ and $\Xi /K$ ratios from central collisions of $%
^{197}$Au + $^{197}$Au at $b=1$ fm.}
\label{aue}
\end{figure}

It is also of interest to study $\Xi $ production from other collision
systems and at different energies. Shown in Fig. \ref{aue}(a) are excitation
functions for the abundances of $K$, $\bar{K}$, and $\Xi $ from central
collisions of $^{197}$Au + $^{197}$Au at $b=1$ \textrm{fm}. The $\Xi $
abundance is about $5.4\times 10^{-6}$, $1.8\times 10^{-4}$, and $9.6\times
10^{-4}$ at $E/A=0.96$, $1.48$, and $2.0$ \textrm{GeV}, respectively.
Therefore, the $\Xi $ yield depends on the incident energy and mass of the
colliding system. Figs. \ref{aue}(b) shows the $\bar{K}/K$ and $\Xi /K$
ratios as functions of incident energy in collisions of $^{197}$Au + $^{197}$%
Au at $b=1$ \textrm{fm}. It is seen that the $\bar{K}/K$ and $\Xi /K$ ratios
increase with increasing incident energy. In particular, large $\Xi $
abundance of about $1.0\times 10^{-3}$ and $\Xi /K$ ratio of about $%
7.3\times 10^{-4}$ are obtained in central $^{197}$Au + $^{197}$Au
collisions at $E/A=2.0$ \textrm{GeV}, which can be studied at AGS \cite%
{e866e917,e895}.

\section{summary}

In summary, we have studied subthreshold $\Xi $ production in heavy-ion
collisions at \textrm{SIS} energies within the framework of a relativistic
transport model that includes explicitly the nucleon, $\Delta$, and pion,
and perturbatively the kaon, antikaon, and hyperons. The production of $\Xi$
is also treated perturbatively through the strangeness-exchange reactions $%
\bar{K}\Lambda \rightarrow \pi \Xi $ and $\bar{K}\Sigma \rightarrow \pi \Xi $
with their cross sections taken from the predictions from a hadronic model.
We find that the $\Xi $ yield is about $10^{-4}$ per event in central
collisions of $^{58}$Ni + $^{58}$Ni at $E/A=1.93$ \textrm{GeV} and can reach
to about $10^{-3}$ per event in central collisions of $^{197}$Au + $^{197}$%
Au at $E/A=2.0$ \textrm{GeV}. The $\Xi $ yield is further found to be more
sensitive to the cross sections for the strangeness-exchange reactions than
to the kaon medium effects. Study of subthreshold $\Xi$ production in heavy
ion collisions at \textrm{SIS} energies thus provides the possibility of
studying the strangeness-exchange reactions and testing the hadronic model
used for evaluating their cross sections.

\begin{acknowledgments}
This paper was based on work supported by the U.S. National Science
Foundation under Grant No. PHY-0098805 and the Welch Foundation under Grant
No. A-1358. L.W.C. was also supported by the National Natural Science
Foundation of China under Grant No. 10105008. Y.T. would like to express his
gratitude to the Cyclotron Institute at TAMU for the warm hospitality during
his visit, and his work was supported by Taiwan's National Science Council
Grant NSC92-2112-M001-64
\end{acknowledgments}

\end{document}